\documentclass[journal=jpcbfk,manuscript=article]{achemso}
\usepackage{achemso}
\usepackage{bm,textcomp}
\usepackage{amsmath} 
\usepackage{amssymb} 
\usepackage{graphicx}
\usepackage{mathtools}
\usepackage{natbib}

\author{Colin D. Kinz-Thompson} 
\affiliation{Department of Chemistry, Columbia University, New York, New York}
\altaffiliation{These authors contributed equally}
\author{Ajeet K. Sharma}
\affiliation{Department of Physics, Indian Institute of Technology, Kanpur, India}
\altaffiliation{These authors contributed equally}
\author{Joachim Frank} 
\affiliation{Howard Hughes Medical Institute, Department of Biochemistry and Molecular Biophysics, Columbia University, New York, New York}
\alsoaffiliation{Department of Biology, Columbia University, New York, New York}
\author{Ruben L. Gonzalez, Jr.}
\affiliation{Department of Chemistry, Columbia University, New York, New York}
\author{Debashish Chowdhury}
\affiliation{Department of Physics, Indian Institute of Technology, Kanpur, India}
\altaffiliation{Corresponding author}
\email{debch@iitk.ac.in}

\bibstyle{achemso}

\title{Quantitative Connection Between Ensemble Thermodynamics and Single-Molecule Kinetics: A Case Study Using Cryogenic Electron Microscopy and Single-Molecule Fluorescence Resonance Energy Transfer Investigations of the Ribosome}


\SectionNumbersOn

\begin{document}

\begin{abstract}
At equilibrium, thermodynamic and kinetic information can be extracted from biomolecular energy landscapes by many techniques. However, while static, ensemble techniques yield thermodynamic data, often only dynamic, single-molecule techniques can yield the kinetic data that describes transition-state energy barriers. Here we present a generalized framework based upon dwell-time distributions that can be used to connect such static, ensemble techniques with dynamic, single-molecule techniques, and thus characterize energy landscapes to greater resolutions. We demonstrate the utility of this framework by applying it to cryogenic electron microscopy (cryo-EM) and single-molecule fluorescence resonance energy transfer (smFRET) studies of the bacterial ribosomal pre-translocation complex. Among other benefits, application of this framework to these data explains why two transient, intermediate conformations of the pre-translocation complex, which are observed in a cryo-EM  study, may not be observed in several smFRET studies.
\end{abstract}
%%%%%%%%%%%%%%%%%%%%%%%%%%%%%%

%%%%%%%%%%%%%%%%%%%%%%%%%%%%%%%%%
\centerline{\bf Graphical Abstract}

%%%%%%%%%%%%%%%%%%%%%%%%%%%%%%%
\begin{figure}[h]
	\centering{\includegraphics{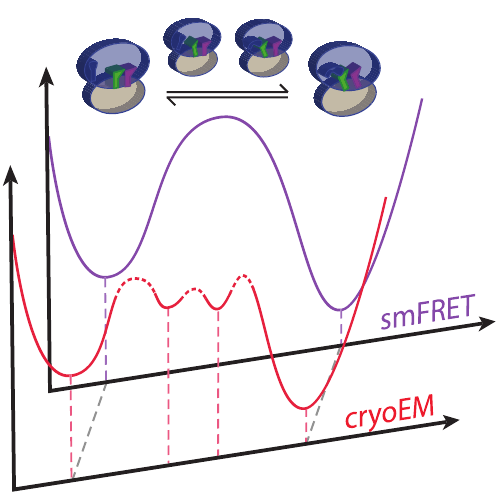}}
%	\centering{\includegraphics{./pic/Fig1_Mechanism.pdf}}
\end{figure}
%%%%%%%%%%%%%%%%%%%%%%%%%%%%%%%%
\newpage
%%%%%%%%%%%%%%%%%%%%%%%%%%%%%%%
\section{INTRODUCTION}

Biomolecular machines operate on energy landscapes with transition-state energy barriers which range from $\sim k_B T$ to the energy of covalent bonds.\cite{Astumian1997,Alberts1998,Peskin1993}
Characterizing the wells and barriers which comprise these energy landscapes is important for understanding the thermodynamics and kinetics of biomolecular machines, and how these thermodynamics and kinetics can be modulated in order to regulate the activities of these machines.\citep{Frank2011,Bustamante2008,Chowdhury2011,Chowdhury2013,Kolomeisky2013,Tinoco2009}
However, due to the stochasticity inherent to these processes, as well as the transient and/or rare nature of states separated by low transition-state energy barriers, extremely sensitive techniques are often required to obtain the level of detail necessary to adequately describe such systems.\citep{Tinoco2011b,Xie2001}
Sufficiently sensitive ensemble techniques, such as cryo-electron microscopy (cryo-EM), can measure static, equilibrium-state populations, and this provides information on the relative energy differences between distinct states.
However, because of the vanishingly small probability of observing a transition state, techniques such as cryo-EM are not able to characterize the transition-state energy barriers responsible for much of the regulation of biomolecular processes.\citep{Frank2013}
Fortunately, dynamic, time-dependent, single-molecule techniques, such as single-molecule fluorescence resonance energy transfer (smFRET), can directly monitor the kinetics of these processes, and allow the characterization of the transition-state energy barriers with theories such as transition-state theory or Kramers' theory.\citep{Zwanzig2001}
smFRET is a particularly powerful technique for connecting single-molecule kinetics to ensemble thermodynamics obtained from cryo-EM in that the FRET efficiency (E$_{\text{FRET}}$) obtained from the smFRET experiments can be correlated to structures obtained from the cryo-EM experiments.
Despite this significant advantage over many other single-molecule techniques, like all techniques, smFRET approaches often suffer from limitations to spatial and temporal resolution, and also often require structural information to develop biologically informative signals.\citep{Kinz-Thompson2014}
Therefore, for any particular system, static, equilibrium-state, ensemble techniques and dynamic, time-dependent, single-molecule techniques provide complementary approaches for studying the underlying biological processes. Nonetheless, given the current limitations in their application, the pictures they provide may not always be congruous. 

The bacterial ribosome is one example of a biological system that has been well studied by both ensemble and single-molecule techniques, although the associated energy landscape remains only coarsely defined.\cite{Frank2010,Munro2009}
Responsible for translating messenger RNAs (mRNAs) into their encoded proteins, the ribosome is composed of a large and small subunit (50S and 30S in bacteria, respectively).
During the elongation stage of translation, the ribosome undergoes consecutive rounds of an elongation cycle in which it successively adds amino acids to the nascent polypeptide chain in the order dictated by the sequence of the mRNA.
In the first step of the elongation cycle, the mRNA-encoded aminoacyl-transfer RNA (aa-tRNA) is delivered to the aa-tRNA binding (A) site of the ribosome in the form of a ternary complex (TC) that is composed of the ribosomal guanosine triphosphatase (GTPase) elongation factor (EF) Tu, guanosine triphosphate (GTP), and aa-tRNA.\citep{Schmeing2009,Voorhees2013,Johansson2008,Rodnina2012}
Upon delivery of the mRNA-encoded aa-tRNA into the A site, peptide bond formation results in transfer of the nascent polypeptide chain from the peptidyl-tRNA bound at the ribosomal peptidyl-tRNA binding (P) site to the aa-tRNA at the A site, generating a ribosomal pre-translocation (PRE) complex carrying a newly deacylated tRNA at the P site and a newly formed peptidyl-tRNA, extended by one amino acid, at the A site.\citep{Rodnina2011,Simonovic2009,Beringer2007}
Subsequently, the ribosome must translocate along the mRNA, moving the newly deacylated tRNA from the P site to the ribosomal tRNA exit (E) site and the newly formed peptidyl-tRNA from the A site to the P site.\citep{Moazed1989,Frank2010,Voorhees2013,Chen2012b,Rodnina2011a,Cooperman2011,Noller2011,Munro2010,Shoji2009}
 While  translocation can occur spontaneously, albeit slowly, \textit{in vitro} \citep{Gavrilova1976}, it is accelerated by orders of magnitude \textit{in vivo} through the action of EF-G, another ribosomal GTPase.\citep{Shoji2009,Voorhees2013}

\begin{figure}[t!]
	\centering{\includegraphics{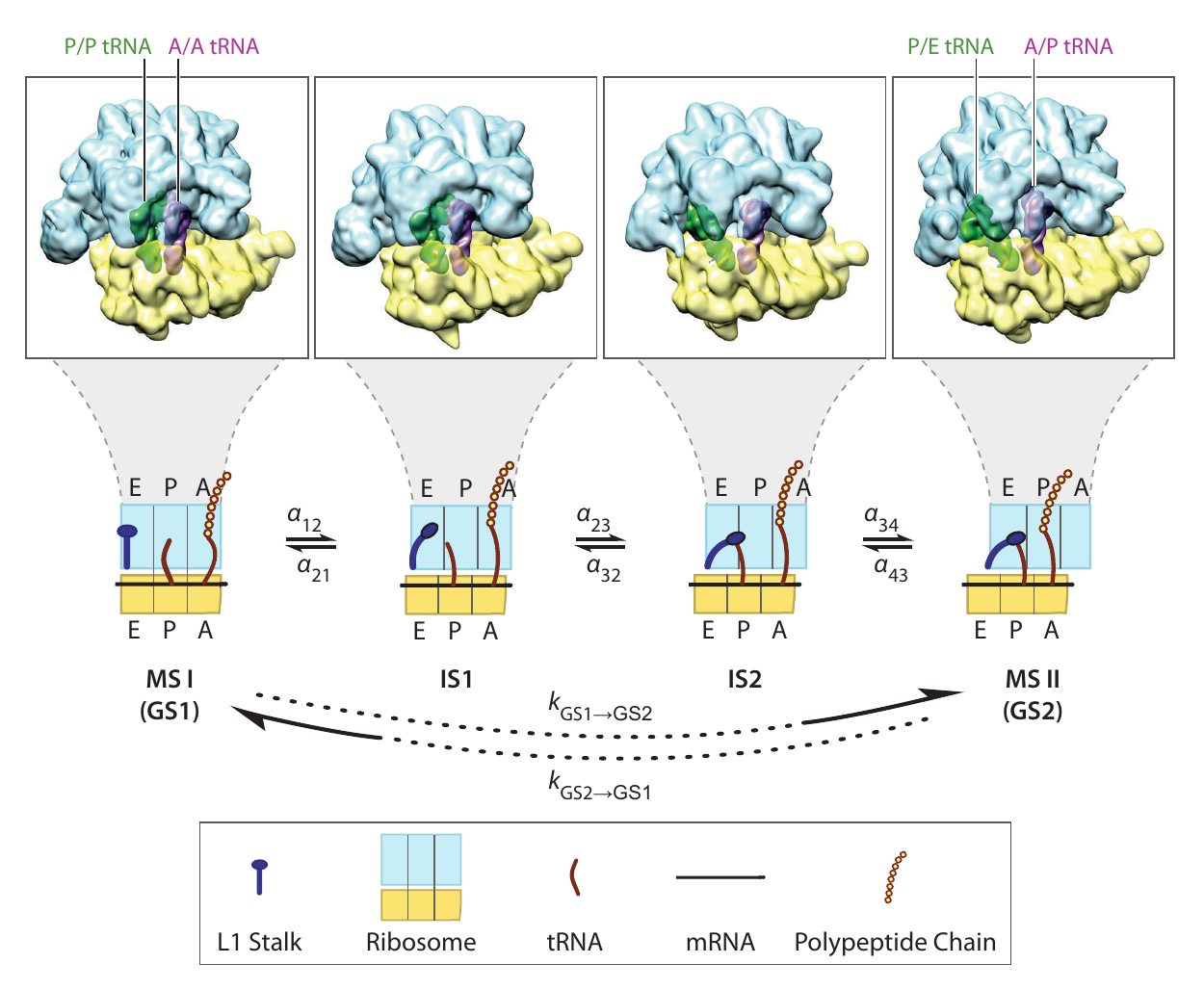}}
%	\centering{\includegraphics{./pic/Fig1_Mechanism.pdf}}
	\caption{Cartoon schematic mechanism of PRE complex fluctuations. After peptide-bond formation, the PRE complex fluctuates between MS I/GS1 and MS II/GS2, passing through IS1 and IS2, until EF-G catalyzed translocation occurs.}
	\label{fig:mechanism}
\end{figure}

Prior to translocation and in the absence of EF-G, at least three individual structural elements of the PRE complex undergo thermally driven conformational fluctuations: 
(i) the P- and A-site tRNAs fluctuate between their classical P/P and A/A configurations and their hybrid P/E and A/P configurations (where, relative to the classical P/P and A/A configurations, the hybrid P/E and A/P configurations are characterized by the movement of the acyl acceptor ends of the P- and A-site tRNAs from the P and A sites of the 50S subunit into the E and P sites of the 50S subunit, respectively);
(ii) the ribosome fluctuates between its non-rotated and rotated subunit orientations (where, relative to the non-rotated subunit orientation, the rotated subunit orientation is characterized by a counterclockwise rotation of the 30S subunit relative to the 50S subunit when viewed from the solvent-accessible side of the 30S subunit);\citep{Frank2000} and
(iii) the L1 stalk of the 50S subunit fluctuates between its open and closed conformations (where, relative to the open L1 stalk conformation, the closed L1 stalk conformation is characterized by movement of the L1 stalk into the intersubunit space such that it can make a direct contact with the hybrid P/E-configured tRNA) (Fig. \ref{fig:mechanism}).\citep{Frank2012}
Because of the stochastic nature of thermally driven processes, the tRNAs, ribosomal subunits, and L1 stalk within an ensemble of PRE complexes will asynchronously fluctuate between these transiently populated states in the absence of EF-G. While this structural heterogeneity impedes ensemble studies of these dynamics, they have been successfully characterized by single-molecule methods.\citep{Fei2008,Fei2009,Blanchard2004c,Cornish2008,Cornish2009,Wang2007a,Chen2011,Kim2007,
Munro2007,Munro2010,Wang2011a,Ning2014}

%%%%%%%%%%%%%%%%%%%%%%%%%%%%%%%
%%%%%%%%%%%%%%%%%%%%%%%%%%%%%%%%

Remarkably, smFRET studies performed by Fei and coworkers have observed PRE complexes fluctuating between two discrete states: 
(i) global state 1 (GS1), characterized by classically configured tRNAs, non-rotated subunits, and an open L1 stalk, and 
(ii) global state 2 (GS2), characterized by hybrid-configured tRNAs, rotated subunits, and a closed L1 stalk. \citep{Fei2008,Fei2009} The observation that the PRE complex fluctuates between just two states in the smFRET studies of Fei and coworkers is consistent with numerous subsequent smFRET studies from several other groups in which the tRNAs, ribosomal subunits, or L1 stalk elements of PRE complexes are also observed to fluctuate between just two states corresponding to the classical and hybrid tRNA configurations, the non-rotated and rotated subunit orientations, or the open and closed L1-stalk conformations, respectively.\citep{Fei2008, Fei2009, Fei2011, Ning2014, Cornish2008, Cornish2009, Chen2011} Furthermore, the initial observation by Fei and coworkers that the PRE complex fluctuates between GS1 and GS2 is consistent with the more recent observation that fluctuations of the tRNAs between their classical and hybrid configurations, the ribosomal subunits between their non-rotated and rotated orientations, and the L1 stalk between its open and closed conformations are physically coupled, and coordinated by the ribosome in order to maximize and regulate the efficiency of translocation.\citep{Ning2014} smFRET studies reveal that the thermodynamics and kinetics of the equilibrium between GS1 and GS2 are sensitive to: (i) the presence, identity, and acylation status of the P-site tRNA;\citep{Fei2008,Fei2009,Blanchard2004c,Cornish2008,Cornish2009,Kim2007,Munro2007,Munro2010,Wang2007a,Chen2011} (ii) the presence and acylation status of the A-site tRNA;\citep{Fei2008,Fei2009,Blanchard2004c,Cornish2008,Cornish2009,Kim2007,Munro2007,Munro2010,Wang2007a,Chen2011} (iii) the binding of EF-G;\citep{Fei2008,Fei2009,Cornish2008,Cornish2009,Wang2007a,Chen2011} (iv) Mg$^{2+}$ concentration;\citep{Kim2007} (v) temperature;\citep{Wang2011a} (vi) the binding of ribosome-targeting antibiotic inhibitors of translocation;\citep{Kim2007,Wang2012b,Ning2014} and (vii) the perturbation of intersubunit rotation via disruption of specific ribosomal intersubunit interactions.\citep{Munro2007,Ning2014} Collectively, these studies have provided deep insights into the roles that the P- and A-site tRNAs, EF-G, antibiotics, cooperative conformational changes, and allostery play in regulating translocation. 

Cryo-EM studies of PRE complexes performed by Agirrezabala and coworkers have directly observed two states that are presumably the structural equivalents of GS1 and GS2, termed macrostate 1 (MS I) and macrostate 2 (MS II), respectively.\citep{Frank2000,Valle2003,Frank2007,Agirrezabala2008} More recent molecular dynamics simulations of cryo-EM-derived structural models of MS I and MS II support this presumption, and found that, in the studies of Fei and coworkers, the E$_{\text{FRET}}$ observed in GS1 and GS2 are consistent with the simulations of MS I and MS II, respectively.\citep{Trabuco2010} However, in addition to  MS I and MS II, a more recent analysis of the  cryo-EM data set of Agirrezabala \textit{et al.} has revealed the presence of two additional states that are presumably intermediate between MS I/GS1 and MS II/GS2 along the reaction coordinate.\citep{Agirrezabala2012} Neither of these two intermediate states, referred to here as intermediate state 1 (IS1) and intermediate state 2 (IS2), nor any others, were detected in the smFRET studies of Fei and coworkers\citep{Fei2008,Fei2009,Fei2011,Ning2014} or several other groups.\citep{Cornish2008, Cornish2009, Chen2011} In contrast, we note that smFRET studies of PRE complexes by Munro and coworkers identified and characterized two additional states that presumably lie along the reaction coordinate between MS I/GS1 and MS II/GS2.\citep{Munro2007,Munro2010} However, these studies relied heavily on the use of smFRET data collected using ribosomes in which a substitution mutation disrupts a critical ribosome-tRNA interaction, and consequently causes the P-site tRNA in the resulting intermediate states to adopt conformations that are very different from those observed in either IS1 or IS2 in PRE complexes formed using wild-type ribosomes.\citep{Agirrezabala2012}

Since the smFRET experiments of Fei and coworkers and the cryo-EM experiments of Agirrezabala and coworkers interrogate PRE complexes composed of wild-type ribosomes (\textit{i.e.,} without mutations that disrupt ribosome-tRNA interactions), it is highly likely that IS1 and IS2 are also present in the smFRET data, but that, given the spatial resolution, time resolution, and/or signal-to-background ratio (SBR) of the smFRET experiments, IS1 and IS2 do not produce large enough changes in E$_{\text{FRET}}$ to be distinguished from GS1 or GS2. By connecting the results from static, equilibrium-state, ensemble experiments, such as cryo-EM, with the results from dynamic, time-dependent, single-molecule experiments, such as smFRET, through a theoretical framework, these hypotheses can be tested, and the energy landscape where the PRE complex exists can be characterized more precisely. Here, we present such a framework based upon equilibrium-state probabilities and dwell-time distributions (see Refs. \citenum{Garai2009a} and \citenum{Sharma2011}, and references therein). This framework is general, and can be applied to various other ensemble and single-molecule techniques; we use a linear kinetic model, but emphasize that the equations can also be derived for other models. As an illustrative case study, we apply this generalized framework to analyze the data obtained from the  cryo-EM and smFRET studies of Agirrezabala \textit{et al.} and Fei \textit{et al.}, respectively. In doing so, we connect the distribution of the MS I, MS II, IS1, and IS2 states of the PRE complex observed by cryo-EM to the transition rates observed between the GS1 and GS2 states observed by smFRET.

%%%%%%%%%%%%%%%%%%%%%%%%%%%%%%%
\section{METHODS}
%%%%%%%%%%%%%%%%%%%%%%%%%%%%%%%%%

%%%%%%%%%%%%%%%%%%%%%%%%%%%%%%%%
\subsection{Dwell-Time Distribution Framework for N-state Markov chain}
%%%%%%%%%%%%%%%%%%%%%%%%%%%%%%%%

Consider two distinct chemical states, 1 and N, of the system connected linearly by a number of on-pathway intermediate states, 2 through N-1, with transitions from state i to state j occurring at rate $\alpha_{ij}$,
\begin{equation}
1 \xrightleftharpoons[\alpha_{21}]{\alpha_{12}} 2 \xrightleftharpoons[\alpha_{32}]{\alpha_{23}} \cdots \xrightleftharpoons[\alpha_{N-1,N-2}]{\alpha_{N-2,N-1}} N-1 \xrightleftharpoons[\alpha_{N,N-1}]{\alpha_{N-1,N}}N.
\label{rxn:n-dim_scheme}
\end{equation}

If $P_{\mu}(t)$ is the probability of finding the system in chemical state $\mu$ at time $t$, then the time evolution of these probabilities are governed by a set of coupled master equations. The steady-state of this set of equations, corresponding to the the constraints $\partial P_{\mu} / \partial t = 0$, yields the equilibrium-state occupation probabilities $P_{\mu}^{eq}$ of populating each state. The distribution of times taken by the system to reach one terminus from the other can be calculated by modifying the original kinetic scheme imposing an absorbing boundary at the destination state. If the destination state is N, then the corresponding modified kinetic scheme would be,

\begin{equation}
1 \xrightleftharpoons[\alpha_{21}]{\alpha_{12}} 2 \xrightleftharpoons[\alpha_{32}]{\alpha_{23}} \cdots \xrightleftharpoons[\alpha_{N-1,N-2}]{\alpha_{N-2,N-1}} N-1 \xrightarrow[]{\alpha_{N-1,N}} N.
\label{rxn:n-dim_firstpassage}
\end{equation} 

By writing master equations for this new scheme, and adopting the method of Laplace transform, we analytically calculated the probability, $f^{p}(t) dt$, that the system, initially at state 1, reaches state N in the time interval between $t$ and $t+dt$. By evaluating the first moment of this distribution, the mean time for this transition, $\langle t_p \rangle$, can be calculated. An analogous process can be performed to calculate the probability, $f^{r}(t) dt$, that the system, initially in state N, reaches state 1 in the time interval between $t$ and $t+dt$, and hence obtain the mean time for this transition, $\langle t_r \rangle$.

The two expressions for the mean transition times between the termini, $\langle t_p \rangle$ and $\langle t_r \rangle$, form a system of equations with all 2N-2 $\alpha_{ij}$ as variables. Ratios of the  $P_{\mu}^{eq}$ define relationships between the rate constants $\alpha_{ij}$; so, if the equilibrium-state probabilities are known, substitution of these ratios into the expressions for $\langle t_p \rangle$ and $\langle t_r \rangle$ reduces the number of degrees of freedom in the system of equations. With an experimental measure of the mean transition time between the terminal states, the system of equations can be solved for the $\alpha_{ij}$ rate constants. The four-state model (two intermediate states) is solved in Appendix A, as is the derivation of the expression for the variances of $t_p$ and $t_r$. Equivalent expressions for the three-state model (one intermediate state) are in Appendix B.

%%%%%%%%%%%%%%%%%%%%%%%%
\subsection{smFRET Simulations}
%%%%%%%%%%%%%%%%%%%%%%%

We simulated 100 E$_{\text{FRET}}$ versus time trajectories with a linear, three-state kinetic scheme. Dwell-times prior to transitions to other states were exponentially distributed according to the appropriate rate constants. In each E$_{\text{FRET}}$ versus time trajectory, the value of $E_{\text{FRET}}$ corresponding to each state was randomized by choosing $r$ from a normal distribution, and calculating $E_{\text{FRET}} = (1+\left(r/R_0\right)^6)^{-1}$, where $R_0$ is the F\"orster radius. For each E$_{\text{FRET}}$ versus time trajectory, $R_0$ was also randomly chosen from a normal distribution. Noise reflecting a reasonable SBR for the total-internal reflection fluorescence (TIRF) microscope used in the smFRET experiments (\emph{i.e.}, $\sigma=0.05$) was also added to each E$_{\text{FRET}}$ versus time trajectory. More details can be found in Appendix C.

%%%%%%%%%%%%%%%%%%%%%%%%%%
\section{RESULTS AND DISCUSSION}
%%%%%%%%%%%%%%%%%%%%%%%%%%%

%%%%%%%%%%%%%%%%%%%%%%%%%%%%%%%%
\subsection{General Framework}\label{sec:gen_framework}
%%%%%%%%%%%%%%%%%%%%%%%%%%%%%%%%

Because of their complexity, biomolecular systems are often investigated with multiple techniques -- each with their individual strengths and weaknesses.
However, different techniques occasionally yield disparate mechanistic pictures that must ultimately be resolved.
One situation in which this problem manifests itself is when a static, equilibrium-state technique such as cryo-EM detects on-pathway intermediates, but a dynamic, time-dependent technique such as smFRET does not.
This situation could arise if the dynamic technique is not sensitive enough to distinguish the intermediate state from other states of the biomolecular system.
In order to reconcile such contrasting measurements, we need to estimate the lifetime of the transient intermediates if these, indeed, exist.
In an effort to get these estimates we consider a linear kinetic pathway with on-pathway intermediates, such as in equation \ref{rxn:n-dim_scheme}, though the framework presented here can easily be extended to include off-pathway intermediates.
Note that for an N-state linear kinetic scheme there are 2N-2 rate constants $\alpha_{ij}$.
Therefore, in principle, the numerical values of all the individual $\alpha_{ij}$ could be obtained if 2N-2 independent algebraic equations satisfied by these rate constants were available.
As we argue now, except for some small values of N, the rate constants $\alpha_{ij}$ are usually underdetermined by the available experimental information. 

Time-dependent smFRET experiments are typically analyzed with a hidden Markov model (HMM).\citep{Kinz-Thompson2014,Bronson2009,Bronson2010,VandeMeent2013,VandeMeent2014} 
Among other things, such an analysis yields HMM-idealized state versus time trajectories from which a distribution of lifetimes in a particular state can be calculated.
If sufficiently transient, on-pathway intermediates between the initial and final state exist, the distribution of idealized lifetimes will not appear significantly different from what it would be in the absence of the intermediate states (\textit{e.g.,} an exponential distribution for a random transition with a time-independent probability of occurrence).
In such a case, the simplest model that the  smFRET data supports is that of a transition with no intermediate states; so, assuming Markovian transitions, the mean lifetimes obtained from the HMM-idealized trajectories would be taken to be the inverses of the effective rate constants for the transitions between the initial and final states.
In contrast, the analytical expressions for the mean lifetimes spent traveling between the terminal states, via intermediate states, of a N-state kinetic scheme contain the rate constants $\alpha_{ij}$ that describe the direct transitions between the intermediate states (see Appendices A and B).
Therefore, equating the mean lifetimes for the forward and reverse transitions between the terminal states that were inferred from dynamic, single-molecule experiments with the corresponding theoretically calculated mean lifetimes yields two algebraic equations that involve 2N-2 rate constants $\alpha_{ij}$.

Thus, in practice, except for the trivial case of N=2, the information available in the form of the effective rates of forward and reverse transitions between the two terminal states would be inadequate to determine all the 2N-2 rate constants $\alpha_{ij}$  that describe the full kinetic mechanism. Obviously, for larger values of N, the number of degrees of freedom must be reduced further by acquiring additional experimental information. This extra information comes from the equilibrium-state experiments. Including information about the equilibrium-state probabilities for the N states provides N-1 additional independent equations (the constraint of normalization of the probabilities, i.e., their sum must be equal to unity, reduces the number from N to N-1). So, for N=3 one would have just enough information to write down four independent equations satisfied by the four rate constants in the three-state model. However, for N=4, we have fewer equations than the number of unkowns and, therefore, in the absence of any other information, one of the rate constants would remain a free parameter. Any one of the six rate constants can be selected as the free parameter. Then, as we will show later in this section, varying the selected free parameter allows one to enumerate all the solutions which are consistent with the data, and thereby impose lower and/or upper bounds on the magnitudes of the rates. In case of higher values of N, the analytical expressions for the variance of $t_p$ and $t_r$, reported in the Appendix, can be utilized for further reduction of the number of degrees of freedom if the corresponding experimental data becomes available in the future.

%%%%%%%%%%%%%%%%%%%%%%%%%%%%%%%%%%%%%%%%%%%%%%
\subsection{Model System: The Bacterial Pretranslocational Complex} \label{sec:pretranslocation}
%%%%%%%%%%%%%%%%%%%%%%%%%%%%%%%%%%%%%%%%%%%%%%

Agirrezabala and coworkers collected cryo-EM data on PRE complexes containing tRNA$^{fMet}$ in the P site, and fMet-Trp-tRNA$^{Trp}$ in the A site.\citep{Agirrezabala2008} Using ML3D, a maximum-likelihood based classification method, particles from this data set were more recently classified into six classes. \citep{Agirrezabala2012} 
The conclusions of this study strongly suggest that three of the classes represent MS I and MS II--MS II being comprised of two structurally similar classes. Additionally, the authors propose that two of the other classes represent on-pathway intermediate states (IS1 and IS2, respectively) between MS I and MS II. As the remaining class represents PRE complexes that are missing a tRNA in the A site, it is therefore ignored. Thus, the model of PRE dynamics proposed by this study is,
\begin{align}
1 \xrightleftharpoons[\alpha_{21}]{\alpha_{12}} 2 \xrightleftharpoons[\alpha_{32}]{\alpha_{23}} 3 \xrightleftharpoons[\alpha_{43}]{\alpha_{34}} 4,
\label{rxn:4state}
\end{align}
where states 1, 2, 3, and 4 represent MS I, IS1, IS2, and MS II, respectively, and are distributed as shown in table \ref{table:4s_cryo}.

%%%%%%%%%%%%%%%%%%%%%%%%%%%%%%
\begin{table} %[!HTB] \label{table:4s_cryo}
\begin{center} 
\begin{tabular}{c |c |c  | c} 
State Index ($\mu$) & State & Class &  $P_{\mu}^{eq}$ \\
\hline
1 & MS I  & 2 &  0.231 \\
2 & IS1  & 4A &  0.131 \\
3 & IS2  & 4B &  0.140 \\
4 & MS II  & 5/6&  0.498 \\
\end{tabular}
\caption{A summary of the ribosomal PRE complexes observed by Agirrezabala and coworkers.\citep{Agirrezabala2012}} \label{table:4s_cryo}
\end{center}
\end{table}
%%%%%%%%%%%%%%%%%%%%%%%%%%%%%%%

%smFRET Data% 

Similarly, smFRET experiments performed by Fei and coworkers monitored PRE complexes as they transitioned, driven by thermal energy, between two global conformational states, GS1 and GS2, which correspond structurally to MS I and MS II, respectively.\citep{Fei2008,Fei2009} By monitoring the relative change in the distance between the P-site tRNA and the ribosomal protein L1 within the L1 stalk of the 50S subunit, the smFRET signal developed by Fei and coworkers probably reports upon the tRNA motions along the pathway proposed by Agirrezabala and coworkers. However, these smFRET measurements were performed for several PRE complexes of variable composition. Of these complexes, perhaps the most relevant complex to the work performed by Agirrezabala and coworkers is the PRE$_{\text{fM/F}}$ complex carrying a tRNA$^{\text{fMet}}$ in the P site and a fMet-Phe-tRNA$^{\text{Phe}}$, rather than a fMet-Trp-tRNA$^{\text{Trp}}$, in the A site. Since the identity of the A-site dipeptidyl-tRNA in the two experiments differs, this could potentially lead to tRNA-dependent differences in the populations and lifetimes of the various states and, consequently, the rates of transitions between these states. Indeed, smFRET studies have shown that the lifetimes of GS1 and GS2 do depend on the presence\citep{Fei2008,Fei2009} and acylation status (\textit{i.e.}, deacylated tRNA$^{\text{Phe}}$ versus Phe-tRNA$^{\text{Phe}}$ versus fMet-Phe-tRNA$^{\text{Phe}}$) of the A-site dipeptdyl-tRNA.\citep{Blanchard2004c,Kim2007} It should be noted, however, that the effect of the identity of the A-site dipeptidyl-tRNA itself (\textit{i.e.}, tRNAs other than tRNA$^{\text{Phe}}$) has not yet been tested by smFRET. In addition to the difference in the identity of the A-site dipeptidyl-tRNA in the cryo-EM and smFRET studies, the Mg$^{2+}$ concentrations employed in the two studies differ. The cryo-EM studies were performed at $\left[\text{Mg}^{2+}\right] = 3.5$ mM and the smFRET studies were performed at $\left[\text{Mg}^{2+}\right] = 15$ mM. 
Previously, smFRET studies have demonstrated that changes to the Mg$^{2+}$ concentration over this range affects the populations and lifetimes of the GS1 and GS2 states.\citep{Kim2007}
With this in mind, it is likely that the equilibrium-state populations observed in the cryo-EM experiments and the corresponding state occupancies in the smFRET experiments are disparate.
Nonetheless, despite their experimental differences, these cryo-EM and smFRET studies are the most experimentally similar cryo-EM and smFRET studies of wild-type bacterial PRE complexes that have been reported in the literature.
Therefore, as a case study, we have chosen to quantitatively compare these two particular studies in order to demonstrate the application of the general framework developed in Section \ref{sec:gen_framework}.

The transition rates between GS1 and GS2 reported using the PRE$_{\text{fM/F}}$ complex for the L1-tRNA donor-acceptor labeling scheme were $k_{\text{GS1}\rightarrow\text{GS2}} = 2.8 \pm 0.2 s^{-1}$ and $k_{\text{GS2}\rightarrow\text{GS1}} = 3.0 \pm 0.4 s^{-1}$.\citep{Fei2009} Given that no evidence of intermediate states was observed, this suggests that any intermediates states, if they exist, might be very transient relative to the time resolution with which the smFRET data was acquired. Indeed, there is a limitation to the time resolution with which smFRET data can be acquired with the electron-multiplying charge-coupled device (EMCCD) cameras that are typically used as detectors in TIRF microscopy-based smFRET experiments. 
Transitions that are faster than the EMCCD camera's acquisition rate (20 $s^{-1}$ in Fei, et  al.)\citep{Fei2009} result in time averaging of the E$_{\text{FRET}}$ and the recording of a single, artifactual data point that appears at the time-averaged value of the E$_{\text{FRET}}$ between the states involved in the rapid fluctuations. This is a well-documented feature of smFRET data analysis, which we term ``blurring''.\citep{Bronson2009} This effect is further compounded by the fact that current state-of-the-art computational methods used to analyze the smFRET data cannot distinguish between artificial, short-lived (\textit{i.e.}, one data point) ``states'' resulting from blurring and actual, short-lived (\textit{i.e.}, one data point) states resulting from the sampling of true intermediate states.\citep{Bronson2009} With such an analysis, the true molecular states become hidden among the ``blurred'' states.  

Reanalysis of the original PRE$_{\text{fM/F}}$ data using the software-package ebFRET--a state-of-the-art, HMM-based analysis method for smFRET data\citep{VandeMeent2013,VandeMeent2014}--yields a better estimate of the transition rates. This is because ebFRET uniquely enables analysis of the entire ensemble of individual E$_{\text{FRET}}$ versus time trajectories, instead of analyzing them in the traditional, isolated, one-by-one manner. The two-state rates inferred by means of ebFRET are similar to, though perhaps more accurate than, those reported originally by Fei and coworkers:  $k_{\text{GS1}\rightarrow\text{GS2}} = 2.0 \pm 0.2~ s^{-1}$, and $k_{\text{GS2}\rightarrow\text{GS1}} = 2.8 \pm 0.1~ s^{-1}$ (see Table \ref{table:2s_FRET}). Interestingly, application of ebFRET reveals that the smFRET PRE$_{\text{fM/F}}$ data are best described by a five-state model. However, further analysis indicates that the three additional states are probably artifacts of blurring, because they are negligibly populated, have extremely transient lifetimes, and occur at an E$_{\text{FRET}}$ that is in between the E$_{\text{FRET}}$ of the two well-defined states.

%%%%%%%%%%%%%%%%%%%%%%%%%%%%%%
\begin{table} %[!HTB] \label{table:2s_FRET}
\begin{center} 
\begin{tabular}{c | c | c } 
$k_{GS1 \to GS2}$ (s$^{-1}$) & $k_{GS2 \to GS1}$ (s$^{-1}$) & Reference \\
\hline
2.8 $\pm$ 0.2 &  3.0 $\pm$ 0.4 & \ \citenum{Fei2009} \\ 
2.0 $\pm$ 0.2 &  2.8 $\pm$ 0.1 & This work \\
\end{tabular}
\caption{ A summary of the rates of transition between $GS1$ and $GS2$ observed by Fei and coworkers.\citep{Fei2009}} \label{table:2s_FRET}
\end{center}
\end{table}
%%%%%%%%%%%%%%%%%%%%%%%%%%%%%%%

%%%%%%%%%%%%%%%%%%%%%%%%%%%%%%%%%%
\subsection{Four-State Model of PRE Complex Dynamics}
%%%%%%%%%%%%%%%%%%%%%%%%%%%%%%%%%%

\begin{figure}%[!HTB]
	%\centering
	\includegraphics{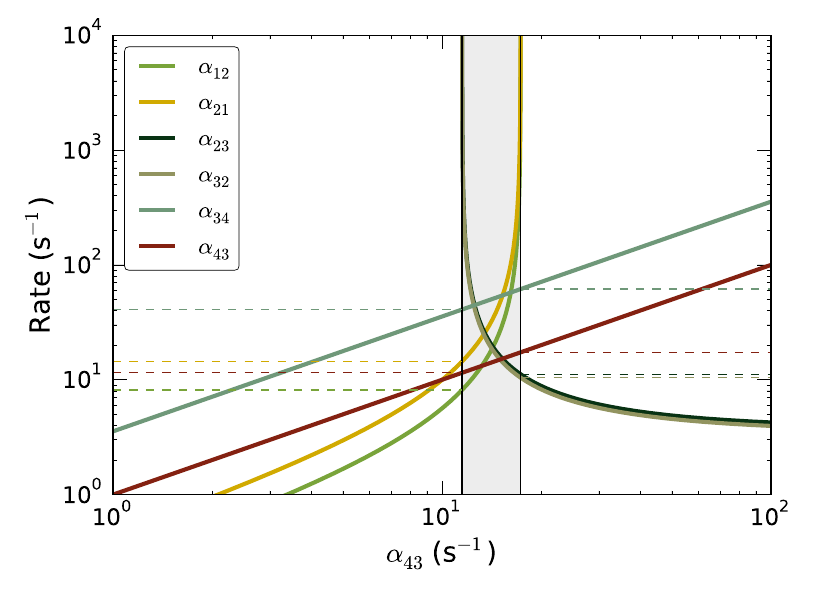}%{./pic/fig_4state_allowed}
	\caption{Rate constants for the four-state model as a function of $\alpha_{43}$. The gray region contains the solutions where all rate constants are positive. Both axes are log scaled; so, any incompatible rates (negative valued) are not shown. The points where rate constants switch from being negative to positive valued are denoted with a black, vertical line. Horizontal, dashed lines denote upper or lower bounds for particular rate constants}
	\label{fig:4state-allowed}
\end{figure}

The dynamics of PRE complexes were analyzed using the general framework presented in section \ref{sec:gen_framework} where the experimental data summarized in Tables \ref{table:4s_cryo} and \ref{table:2s_FRET} were used as constraints for the linear, four-state kinetic scheme shown in Equation \ref{rxn:4state}. For this kinetic scheme, the mean time needed for the forward transition from the terminal state 1 to the terminal state 4, $<t_{p}>$, is given by (see Appendix A for the full derivation)
\begin{equation} \label{eqn:4s-tptext}
<t_{p}>=\dfrac{1}{\alpha_{12}}\biggl[1+\dfrac{\alpha_{21}}{\alpha_{23}}+\dfrac{\alpha_{21}\alpha_{32}}{\alpha_{23}\alpha_{34}}\biggr]+\dfrac{1}{\alpha_{23}}\biggl[1+\dfrac{\alpha_{32}}{\alpha_{34}}\biggr]+\dfrac{1}{\alpha_{34}}.
\end{equation}
The corresponding mean time for the reverse transition from the state 4 to the state 1, $<t_{r}>$, is given by (see Appendix A for the full derivation)
\begin{equation} \label{eqn:4s-trtext}
<t_{r}>=\dfrac{1}{\alpha_{43}}\biggl[1+\dfrac{\alpha_{34}}{\alpha_{32}}+\dfrac{\alpha_{34}\alpha_{23}}{\alpha_{32}\alpha_{21}}\biggr]+\dfrac{1}{\alpha_{32}}\biggl[1+\dfrac{\alpha_{23}}{\alpha_{21}}\biggr]+\dfrac{1}{\alpha_{21}}.
\end{equation}
The probabilities for the occupation of the four states at equilibrium are given by (see Appendix A for the full derivation)
\begin{align} \label{eqn:4s_pst1text}
P_{1}^{eq}&=\frac{\alpha_{43}\alpha_{32}\alpha_{21}}{\alpha_{43}\alpha_{32}\alpha_{21}+\alpha_{12}\alpha_{43}\alpha_{32}+\alpha_{12}\alpha_{23}\alpha_{43}+\alpha_{12}\alpha_{23}\alpha_{34}}, \\ \label{eqn:4s_pst2text}
P_{2}^{eq}&=\frac{\alpha_{12}\alpha_{43}\alpha_{32}}{\alpha_{43}\alpha_{32}\alpha_{21}+\alpha_{12}\alpha_{43}\alpha_{32}+\alpha_{12}\alpha_{23}\alpha_{43}+\alpha_{12}\alpha_{23}\alpha_{34}}, \\ \label{eqn:4s_pst3text}
P_{3}^{eq}&=\frac{\alpha_{12}\alpha_{23}\alpha_{43}}{\alpha_{43}\alpha_{32}\alpha_{21}+\alpha_{12}\alpha_{43}\alpha_{32}+\alpha_{12}\alpha_{23}\alpha_{43}+\alpha_{12}\alpha_{23}\alpha_{34}},\text{ and} \\ \label{eqn:4s_pst4text}
P_{4}^{eq}&=\frac{\alpha_{12}\alpha_{23}\alpha_{34}}{\alpha_{43}\alpha_{32}\alpha_{21}+\alpha_{12}\alpha_{43}\alpha_{32}+\alpha_{12}\alpha_{23}\alpha_{43}+\alpha_{12}\alpha_{23}\alpha_{34}},
\end{align}
with the normalization condition $\sum_{\mu=1}^4 P_{\mu}^{eq} = 1$.
Using the equations for $<t_{p}>$ and $<t_{r}>$ (Equations \ref{eqn:4s-tptext} and \ref{eqn:4s-trtext}, respectively), and the equations for $P^{eq}_{\mu}$ (Equations \ref{eqn:4s_pst1text}-\ref{eqn:4s_pst4text}), a plot was generated of all the rate constants $\alpha_{ij}$ as functions of an independent $\alpha_{43}$ (Fig. \ref{fig:4state-allowed}). 

Notably, for some values of the independent rate constant, solutions for the dependent rate constants are negative. While this is a consistent solution of the model, only the values where all rate constants are positive are physically-relevant solutions. The boundaries to this region where all rate constants are positive therefore represent the upper or lower bounds on the rate constants for the four-state model that are consistent with both the cryo-EM and the smFRET studies.

%%%%%%%%%%%%%%%%%%%%%%%%%%%%%%%
%%%%%%%%%%%%%%%%%%%%%%%%%%%%%%%%

Interestingly, Fig. \ref{fig:4state-allowed} depicts the upper and lower bounds for $\alpha_{34}$ and $\alpha_{43}$, but only lower-bound cutoffs for the other $\alpha$'s. These other rate constants all asymptotically converge to positive infinity; $\alpha_{12}$ and $\alpha_{21}$ increase with increasing $\alpha_{43}$, while $\alpha_{23}$ and $\alpha_{32}$ compensate by decreasing to their lower-bound. Plotting these results as a function of an independent $\alpha_{12}$ or $\alpha_{23}$ yields the same bounds--there is only a narrow window where all $\alpha$'s are consistent with the cryo-EM and smFRET data. With such boundaries on the individual rate constants, one can estimate the EMCCD camera acquisition rate that would be needed to distinctly observe the transient states of interest.

%%%%%%%%%%%%%%%%%%%%%%%%%%%%%%%%%%%%%
\subsection{Three-State Pretranslocation Model} \label{sec:3s-results}
%%%%%%%%%%%%%%%%%%%%%%%%%%%%%%%%%%%%%

Since the number of equations available in our four-state model is five whereas the number of unknown rate constants is six, we could only express five rate constants in terms of the sixth one. In contrast, because we can reduce the number of states from four to three (\textit{vide infra}), in this subsection we use the four corresponding independent equations to extract the absolute values of the four rate constants associated with the three-state model,
\begin{align}
1 \xrightleftharpoons[\alpha_{21}]{\alpha_{12}} 2 \xrightleftharpoons[\alpha_{32}]{\alpha_{23}} 3. 
\label{rxn:3state}
\end{align} 
As explained in Appendix C, structural analysis strongly suggests that the L1-tRNA distance in IS1 is insufficiently different from that of MS I so as to result in an E$_{\text{FRET}}$ that is significantly different than that of MS I. Thus, MS I and IS1 can be combined into a single state, state 1, thereby reducing the four-state model into a three-state kinetic scheme of Eqn. \ref{rxn:3state}, where the states 2 and 3 correspond to IS2 and MS II, respectively.

The expressions for $<t_{p}>$, $<t_{r}>$, and $P_{\mu}$ ($\mu=1,2,3$) are given by (see Appendix B for detailed derivations),
\begin{equation}
<t_{p}>=\dfrac{1}{\alpha_{12}}\biggl[1+\dfrac{\alpha_{21}}{\alpha_{23}}\biggr]+\dfrac{1}{\alpha_{23}},
\end{equation}
\begin{equation}
<t_{r}>=\dfrac{1}{\alpha_{32}}\biggl[1+\dfrac{\alpha_{23}}{\alpha_{21}}\biggr]+\dfrac{1}{\alpha_{21}},
\end{equation}
\begin{align}
P_{1}^{eq}&=\frac{\alpha_{21}\alpha_{32}}{\alpha_{21}\alpha_{32}+\alpha_{12}\alpha_{32}+\alpha_{23}\alpha_{12}}, \\
P_{2}^{eq}&=\dfrac{\alpha_{12}\alpha_{32}}{\alpha_{21}\alpha_{32}+\alpha_{12}\alpha_{32}+\alpha_{23}\alpha_{12}},\text{ and} \\
P_{3}^{eq}&=\dfrac{\alpha_{23}\alpha_{12}}{\alpha_{21}\alpha_{32}+\alpha_{12}\alpha_{32}+\alpha_{23}\alpha_{12}},
\end{align}
where the normalization condition is $\sum_{\mu=1}^{eq} P_{\mu}=1$.
These expressions, together with the corresponding experimental data, are utilized to write down four independent equations. The rate constants computed by solving those four equations are shown in Table \ref{table:3s-rates}.

Interestingly, this calculation suggests that the rate limiting steps for the foward and reverse reactions are IS2 $\rightarrow$ MS II and MS II $\rightarrow$ IS2, respectively, while interconversion between MS I/IS1 and IS2 occur relatively rapidly. These rates for MS I/IS1 to IS2 interconversion are approximately the same or faster than the 20 s$^{-1}$ EMMCD camera acquisition rate of the original smFRET data from Fei \textit{et al.}, and, additionally, the change in the L1-tRNA distance between the MS I/IS1 and IS2 states is relatively small ($\sim 80$ \AA~to $64$ \AA), resulting in a correspondingly small difference in E$_{\text{FRET}}$ ($\sim 0.15$~to $0.40$); so, any separation of MS I/IS1 and IS2 that might have been observed in the smFRET data would likely have been obscured in the HMM analysis process by camera blurring arising from interconversion rates that are similar to the acquisition rate. The fast interconversion and small expected changes in E$_{\text{FRET}}$ suggest that MS I, IS1, and IS2 might have originally been interpreted as a `single', averaged state in the analysis of the smFRET data. 

%%%%%%%%%%%%%%%%%%%%%%%%%%%%%%%
\begin{table}[!bht] %[!HTB] \label{table:3s-rates}
\begin{center}
\begin{tabular}{c | c | c }
$\alpha$ & Max k$_{\text{L1-tRNA}}$ ($s^{-1}$) & Mean k$_{\text{L1-tRNA}} \pm 1 \sigma $ ($s^{-1}$) \\
\hline
$\alpha_{12}$ & 18.1 & 23.3 $\pm$ 22.7\\
$\alpha_{21}$ & 46.8 & 52.5 $\pm$ 33.6\\
$\alpha_{23}$ & 5.90 & 5.89 $\pm$ 0.42\\
$\alpha_{32}$ & 1.66 & 1.66 $\pm$ 0.12 
\end{tabular}
\end{center}
\caption{Rate constants for PRE$_{\text{fM/F}}$ using a linear, three-state kinetic scheme where MS I and IS1 have been combined into the first state. Error from the ebFRET-estimated smFRET-determined rate constants, and the counting error from the cryo-EM study were propagated into distributions of the $\alpha$. These distributions are strictly not normal distributions, although $\alpha_{23}$ and $\alpha_{32}$ are approximately normal.} \label{table:3s-rates}
\end{table}
%%%%%%%%%%%%%%%%%%%%%%%%%%%%%%

%%%%%%%%%%%%%%%%%%%%%%%%%%%%%%%%%%%%
\subsection{Synthetic smFRET Time Series} \label{sec:simulation} 
%%%%%%%%%%%%%%%%%%%%%%%%%%%%%%%%%%%%

To investigate how ebFRET would treat this `single', averaged smFRET state, synthetic time series simulating a three-state PRE$_{\text{fM/F}}$ complex were constructed guided by the analysis above. Estimates for $E_{\text{FRET}}$ were based upon the cryo-EM structures of Agirrezabala and coworkers (see Appendix C), and the kinetic scheme and associated rate constants employed are those in Section \ref{sec:3s-results}. Since the rate constants for the transition between states MS I/IS1 and IS2 are of the same order of magnitude as the frame rate of this simulation, traditional smFRET data analysis of this synthetic data provides insight into whether blurring could have obscured any transient, intermediate states in the data of Fei and coworkers. Typically, such obfuscation begins to manifest when dwell-times in a state of interest approach the same order of magnitude as the EMCCD camera acquisition time, because of errors in estimating the lengths of the dwell-times.\citep{Bronson2009} 

\begin{figure}%[!HTB]
%	\centering
	\includegraphics[]{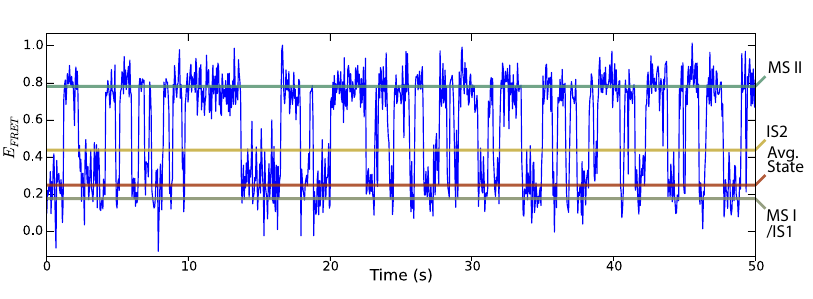}%[height=3.5 in]{./pic/example_synthetics.png}	
	\caption{Example synthetic $E_{\text{FRET}}$ versus time trajectory. The gray, yellow, and blue horizontal lines denote the E$_{\text{FRET}}$ means used to simulate MS I/IS1, IS2, and MS II for this time series, respectively. The red horizontal line is the equilibrium-state-weighted average of the means of the MS I/IS1 and IS2 states.}
	\label{fig:example_timeseries}
\end{figure}

The synthetic smFRET dataset was constructed by carrying out simulations of E$_{\text{FRET}}$ versus time trajectories where each `single-ribosome' had randomized simulation parameters as described in Appendix C. This probabilistic approach accounts for experimental variation (\textit{e.g.,} uneven illumination in the field-of-view), as well as ensemble variations (\textit{e.g.} static disorder from a small sub-population of ribosomes lacking an A-site dipeptidyl-tRNA). An example of a synthetic $E_{\text{FRET}}$ vs. time trajectory is shown in Fig. \ref{fig:example_timeseries}, where the ensemble mean values of E$_{\text{FRET}}$ were $\sim$ 0.16, 0.40, and 0.74 for states MS I/IS1, IS2, and MS II, respectively.

%%%%%%%%%%%%%%%%%%%%%%%%%%%
%%%%%%%%%%%%%%%%%%%%%%%%%%%%

With regard to the distribution of E$_{\text{FRET}}$ values observed from any ensemble of E$_{\text{FRET}}$ versus time trajectories, blurring would result in a shift of some of the density of the equilibrium-state occupancy probability distribution to an intermediate, averaged value between the blurred states. Deviation of a histogram of the observed, simulated E$_{\text{FRET}}$ in the synthetic data set from the distribution predicted by the equilibrium-state occupancies of the linear, three-state model therefore can be used to characterize the amount of blurring present in the synthetic data. We modeled the normalized histogram of the synthetic ensemble with normal distributions weighted by their respective equilibrium-state probability, $P_\mu^{eq}$ (Fig. 4A). The mean of each state was distributed according to the distribution of static E$_{\text{FRET}}$ for that state in each of the synthetic time series (Fig. 4C). This approach accurately reflects a non-blurred histogram of E$_{\text{FRET}}$ (Fig. 5). Deviations that occur are therefore due to blurring, or, if they had been simulated in this synthetic dataset, could have been due to the presence of unaccounted-for states. Notably, a large portion of the MS I/IS1 density in Fig. 4A is relocated into the region between MS I/IS1 and IS2. This is a direct manifestation of blurring. By collapsing MS I/IS1 and IS2 into one averaged state (Fig. 4B and 4C), we find that the data are much better described by only two states (Fig. 4D). In this case, the artifactual, blurred, averaged state overwhelms any distinction between the MS I/IS1 and IS2 states, whereas when the simulation is performed with an acquisition rate that is significantly faster than the transitions of interest (2000 s$^{-1}$), these states are well-resolved (Fig. 5).

\begin{figure}%[!HTB]
%	\centering
	\includegraphics[]{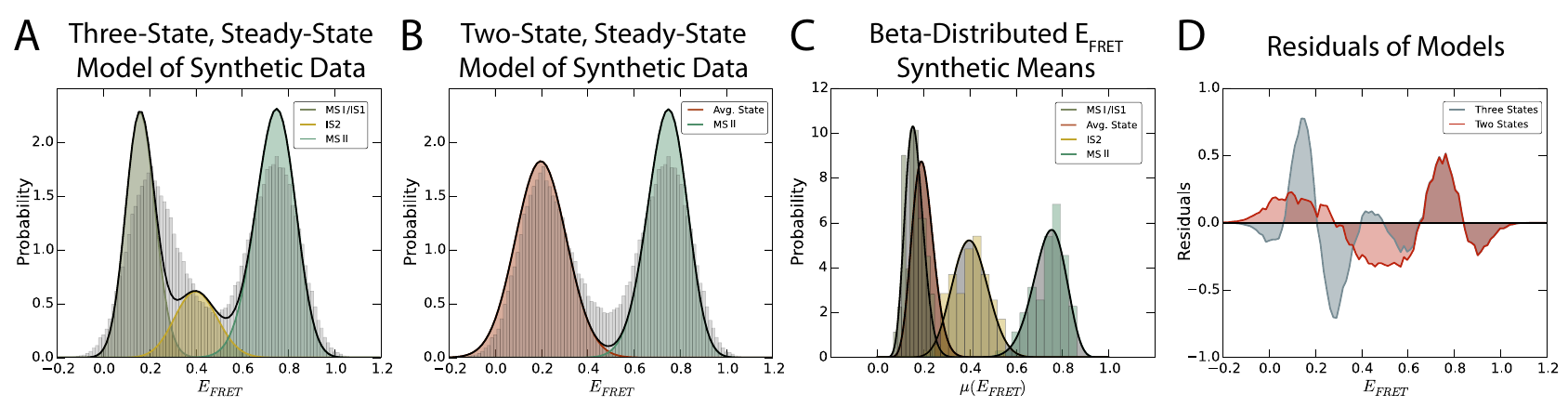}%[height=3.5 in]{./pic/example_synthetics.png}	
	\caption{(A) Histogram of synthetic PRE complex data (gray) modeled with three, non-blurred states. (B) Histogram of synthetic PRE complex data (gray) modeled with two, non-blurred states. The ``Avg. State'' is a weighted combination of MS I/IS1 and IS2. (C) Histograms and probability distributions of the static E$_{\text{FRET}}$ value means generated for the synthetic PRE complex dataset. (D) Plots of the model probability densities minus the normalized histograms from panels A and B. The sum of the squares of these residuals are 10.03, and 4.02 for the three-state and two-state models, respectively, suggesting that the blurred, synthetic, PRE complex data are better represented by a two-state model.}
	\label{fig:distributions}
\end{figure}

\begin{figure}[htb!]
	\centering{\includegraphics{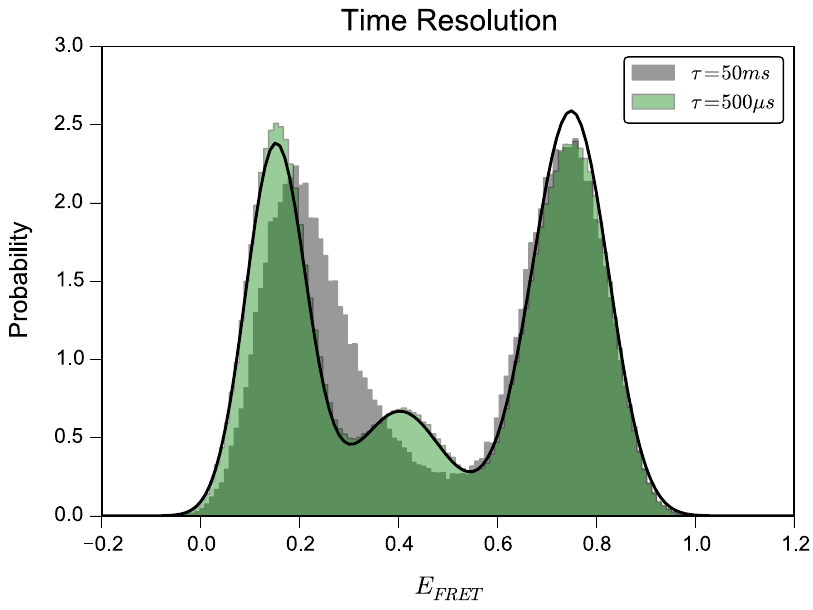}} \\ \begin{flushleft}
%\centering{\includegraphics{./pic/Fig5_TimeResolution.pdf}} \\ \begin{flushleft}
	Figure 5. Histograms of the same ensemble of synthetic E$_{\text{FRET}}$ vs. time trajectories rendered with different time resolutions. The left-most peak in the grey histogram, which was used in Fig. 4, is well-resolved into two separate peaks when the synthetic data is rendered with a 100x faster acquisition rate (green histogram). The three-state model distribution (black curve) is that from Fig. 4, which was modeled using only the initial simulation parameters. Analysis of the faster time resolution time series (green histogram) using ebFRET accurately estimated all four transition rates ($k_{12} = 17.8 \pm 0.6$ s$^{-1}$, $k_{21} = 47.2 \pm 1.6$ s$^{-1}$, $k_{23} = 6.4 \pm 0.6$ s$^{-1}$, $k_{32} = 1.82 \pm 0.17$ s$^{-1}$), as well as accurately estimated the distribution of E$_{\text{FRET}}$ means ($\mu_1 = 0.16$, $\sigma_1 = 0.035$, $\mu_2 = 0.41$, $\sigma_2 = 0.067$, $\mu_3 = 0.74$, $\sigma_3 = 0.058$) and the noise parameter ($\sigma_{noise} = 0.050$).
	\end{flushleft}
\end{figure}

This blurred, synthetic, three-state ensemble of E$_{\text{FRET}}$ versus time trajectories was then analyzed with ebFRET. Interestingly, ebFRET overestimated the true number of kinetic states. Most likely this is due to the fact that the dwell-times in states MS I/IS1 and IS2 are too transient relative to the simulation's `acquisition rate', because ebFRET is able to accurately infer the model parameters from the same data with a more appropriate acquisition rate of 2000 s$^{-1}$ (Fig. 5). As a result, the blurred, synthetic data points are modeled by ebFRET as distinct `states' -- even though these states do not actually exist on the `energy landscape' of the simulated PRE$_{\text{fM/F}}$ complex.

%%%%%%%%%%%%%%%%%%%%%
\section{Conclusion}
%%%%%%%%%%%%%%%%%%%%%

We have established a general, theoretical framework that integrates equilibrium-state cryo-EM observations with dynamic, time-dependent smFRET observations. This framework is not limited exclusively to cryo-EM and smFRET;  any technique that provides static, equilibrium-state populations can be integrated with any other technique that provides dynamic, time-dependent transition rates. The analysis reported here suggests that states IS1 and IS2, previously identified by cryo-EM, are not detected by smFRET because their short lifetimes result in transitions that are too fast to be characterized, given the acquisition rate of the EMCCD camera and the limitations of the manner in which hidden Markov models are implemented. Moreover, structural analyses indicate that the L1-tRNA smFRET signal is unable to yield distinguishable signals for MS I and IS1, given the typical SBR of TIRF microscope-based smFRET measurements. The quantitative analysis of the experimental data, based on the analytical theory presented here, provides a possible explanation for why the PRE complex intermediates observed by cryo-EM (\textit{i.e.}, IS1 and IS2) escaped detection by the smFRET studies of Fei \textit{et al.,} \citep{Fei2008,Fei2009} and, possibly, other groups.\citep{Cornish2008, Cornish2009, Chen2011} Conversely, application of this framework to smFRET data in which intermediate states have been detected, but have been difficult to assign to specific PRE complex structures\citep{Munro2007,Munro2010}, should allow researchers to determine whether and how such intermediates correspond to IS1 and/or IS2. Perhaps more importantly, we can now predict the lifetimes and corresponding rates of transitions into and out of IS1 and IS2 that are crucial for designing future kinetic experiments. This work therefore resolves a discrepancy in the field and opens a path for performing and analyzing future experiments. Furthermore, we hope to use this theoretical framework to make predictions regarding future experiments in which the cryo-EM and smFRET data would be collected under more comparable conditions. This theoretical framework will also be useful in guiding future experimental explorations which include, for example, stabilization of IS1 or IS2 (or any other intermediates that may be ultimately identified) using different tRNAs or mutant ribosomes.%

%%%%%%%%%%%%%%%%%%%%%
\section*{Appendices}
%%%%%%%%%%%%%%%%%%%%%
\setcounter{section}{0}
\renewcommand\thesection{\Alph{section}}

%%%%%%%%%%%%%%%%%%%%%%%%%%%%%%%%%%%%%
\section{Four-State Model}
%%%%%%%%%%%%%%%%%%%%%%%%%%%%%%%%%%%%%

The theoretical results reported here are based on similar approaches followed in Refs. \citenum{Garai2009a} and \citenum{Sharma2011} for analytical calculations of the distribution of the dwell-times of a ribosome.

We use the kinetic scheme
\begin{equation}
1 \mathop{\rightleftharpoons}^{\alpha_{12}}_{\alpha_{21}} 2 \mathop{\rightleftharpoons}^{\alpha_{23}}_{\alpha_{32}} 3 \mathop{\rightleftharpoons}^{\alpha_{34}}_{\alpha_{43}} 4,
\label{scheme}
\end{equation}
where integer indices 1, 2, 3 and 4 represent a discrete chemical state and $\alpha_{ij}$ denotes the transition probability per unit time (\textit{i.e.,} rate constant) for the $i \rightarrow j$ transition. If $P_{\mu}(t)$ is the probability of finding the system in chemical state $\mu$ at time $t$, then the time evolution of these probabilities are governed by the following master equations:
\begin{align}
\frac{dP_{1}(t)}{dt} &= -\alpha_{12}P_{1}(t)+\alpha_{21}P_{2}(t),
\label{dp1} \\
\frac{dP_{2}(t)}{dt} &= \alpha_{12}P_{1}(t)-(\alpha_{23}+\alpha_{21})P_{2}(t)+\alpha_{32}P_{3}(t), \\
\frac{dP_{3}(t)}{dt} &= \alpha_{23}P_{2}(t)-(\alpha_{32}+\alpha_{34})P_{3}(t)+\alpha_{43}P_{4}(t),\text{ and}\\ \label{dp4}
\frac{dP_{4}(t)}{dt} &= \alpha_{34}P_{3}(t)-\alpha_{43}P_{4}(t).
\end{align}
Now we calculate the time-independent occupation probability, $P_{\mu}^{eq}$, of each of these states by finding the equilibrium-state solutions of equation \ref{dp1}-\ref{dp4},
\begin{align} \label{eqn:4s_pst1}
P_{1}^{eq}&=\frac{\alpha_{43}\alpha_{32}\alpha_{21}}{\alpha_{43}\alpha_{32}\alpha_{21}+\alpha_{12}\alpha_{43}\alpha_{32}+\alpha_{12}\alpha_{23}\alpha_{43}+\alpha_{12}\alpha_{23}\alpha_{34}}, \\ \label{eqn:4s_pst2}
P_{2}^{eq}&=\frac{\alpha_{12}\alpha_{43}\alpha_{32}}{\alpha_{43}\alpha_{32}\alpha_{21}+\alpha_{12}\alpha_{43}\alpha_{32}+\alpha_{12}\alpha_{23}\alpha_{43}+\alpha_{12}\alpha_{23}\alpha_{34}}, \\ \label{eqn:4s_pst3}
P_{3}^{eq}&=\frac{\alpha_{12}\alpha_{23}\alpha_{43}}{\alpha_{43}\alpha_{32}\alpha_{21}+\alpha_{12}\alpha_{43}\alpha_{32}+\alpha_{12}\alpha_{23}\alpha_{43}+\alpha_{12}\alpha_{23}\alpha_{34}},\text{ and} \\ \label{eqn:4s_pst4}
P_{4}^{eq}&=\frac{\alpha_{12}\alpha_{23}\alpha_{34}}{\alpha_{43}\alpha_{32}\alpha_{21}+\alpha_{12}\alpha_{43}\alpha_{32}+\alpha_{12}\alpha_{23}\alpha_{43}+\alpha_{12}\alpha_{23}\alpha_{34}}.
\end{align}
We also calculate the distribution of the time spent transitioning from chemical states $1$ to $4$ for the first time by modifying the original kinetic scheme into:
\begin{equation*}
1 \mathop{\rightleftharpoons}^{\alpha_{12}}_{\alpha_{21}} 2 \mathop{\rightleftharpoons}^{\alpha_{23}}_{\alpha_{32}} 3 \mathop{\rightarrow}^{\alpha_{34}} 4
\end{equation*} 
and writing the master equations according to this new scheme:
\begin{align}
\frac{dP_{1}(t)}{dt}&=-\alpha_{12}P_{1}(t)+\alpha_{21}P_{2}(t), \\
\frac{dP_{2}(t)}{dt}&= \alpha_{12}P_{1}(t)-(\alpha_{21}+\alpha_{23})P_{2}(t) +\alpha_{32}P_{3}(t), \\
\frac{dP_{3}(t)}{dt}&= \alpha_{23}P_{2}(t)-(\alpha_{32}+\alpha_{34}) P_{3}(t),\text{ and} \\
\frac{dP_{4}(t)}{dt}&= \alpha_{34} P_{3}(t).
\end{align} 
These equations can be re-written in terms of the following matrix notations:
\begin{equation}
\dfrac{d {\mathbf P}(t)}{dt}={\mathbf M} {\mathbf P}(t),
\label{matrix}
\end{equation}
where $P(t)$ is a column matrix whose elements are $P_{1}(t)$, $P_{2}(t)$ and $P_{3}(t)$, and
\begin{equation}
\textbf{M} =
\begin{bmatrix}
    -\alpha_{12} & \alpha_{21} & 0 \\
    \alpha_{12} & -(\alpha_{21}+\alpha_{23}) & \alpha_{32} \\
    0 & \alpha_{23} & -(\alpha_{34}+\alpha_{32}) \\
   \end{bmatrix}.
\end{equation} 
Now, by introducing the Laplace transform of the probability of kinetic states,
\begin{equation}
\tilde{P_{\mu}}(s)=\int_{0}^{\infty}P_{\mu}(t)e^{-st}dt,
\end{equation}
the solution of equation (\ref{matrix}) in Laplace space is 
\begin{equation}
{\bf \tilde{P}}(s)= (s{\bf I}-{\bf M})^{-1}{\bf P}(0).
\label{laplace}
\end{equation} 
The determinant of matrix $(s{\bf I}-{\bf M})$ is the third-order polynomial
\begin{equation}
(s{\bf I}-{\bf M})^{-1}=a_{3}s^3+a_{2}s^2+a_{1}s+a_{0},
\end{equation}
where
\begin{align}
a_{3}&=1, \\
a_{2}&=\alpha_{12}+\alpha_{21}+\alpha_{23}+\alpha_{32}+\alpha_{34}, \\
a_{1}&=\alpha_{21}\alpha_{34}+\alpha_{21}\alpha_{32}+\alpha_{23}\alpha_{34}+\alpha_{12}\alpha_{23}+\alpha_{12}\alpha_{34}+\alpha_{12}\alpha_{32},\text{ and} \\
a_{0}&=\alpha_{12}\alpha_{23}\alpha_{34}.
\end{align} 
We can solve equation \ref{laplace} by using the initial condition
\begin{equation}
P_{1}(0)=1 ~{\rm and}~ P_{2}(0)= P_{3}(0)= P_{4}(0)= 0.
\label{initial} 
\end{equation} Suppose that the probability of transitioning from chemical state $1$ to $4$ in the time interval of $t$ and $t+\Delta t$ is $f^{p}(t) \Delta t$. Then,
\begin{equation}
f^{p}(t) \Delta t = \Delta P_{4}(t).
\end{equation}
Therefore, we find
\begin{equation}
f^{p}(t) = \dfrac{dP_{4}(t)}{dt}=\omega_{34}P_{3}(t).
\label{3dist} 
\end{equation}
Taking the Laplace transform of equation \ref{3dist} gives
\begin{equation}
{\tilde f^{p}}(s) = \omega_{34}{\tilde P_{3}}(s). 
\label{p3s}
\end{equation}
Now we can calculate an analytical expression for ${\tilde P_{3}}(s)$ by solving Equation \ref{laplace}. Inserting this expression into equation \ref{p3s} yields,
\begin{equation}
{\tilde f^{p}}(s) = \dfrac{\alpha_{12}\alpha_{23}\alpha_{34}}{(s+\omega_{1})(s+\omega_{2})(s+\omega_{3})},
\label{fps}
\end{equation}
where $\omega_{1}$, $\omega_{2}$ and $\omega_{3}$ are the solution of the equation
\begin{equation}
\omega^3- a_{2}\omega^2+a_{1}\omega-a_{0}=0. 
\end{equation}
Taking inverse Laplace transform of Equation \ref{fps} gives 
\begin{equation}
f^{p}(t)=\dfrac{\alpha_{12}\alpha_{23}\alpha_{34}}{(\omega_{1}-\omega_{2})(\omega_{1}-\omega_{3})}e^{-\omega_{1}t}+\dfrac{\alpha_{12}\alpha_{23}\alpha_{34}}{(\omega_{2}-\omega_{1})(\omega_{2}-\omega_{3})}e^{-\omega_{2}t}+\dfrac{\alpha_{12}\alpha_{23}\alpha_{34}}{(\omega_{3}-\omega_{1})(\omega_{3}-\omega_{2})}e^{-\omega_{3}t}.
\end{equation}
Now, solving for the first moment of this distribution,
\begin{equation}
<t_{p}>=\int_{0}^{\infty}t f^{p}(t)dt,
\end{equation}
gives
\begin{equation} \label{eqn:4s-tp}
<t_{p}>=\dfrac{1}{\alpha_{12}}\biggl[1+\dfrac{\alpha_{21}}{\alpha_{23}}+\dfrac{\alpha_{21}\alpha_{32}}{\alpha_{23}\alpha_{34}}\biggr]+\dfrac{1}{\alpha_{23}}\biggl[1+\dfrac{\alpha_{32}}{\alpha_{34}}\biggr]+\dfrac{1}{\alpha_{34}}.
\end{equation}
Similarly, the second moment is
\begin{equation}
<t_{p}^2>=\int_{0}^{\infty}t^2 f^{p}(t)dt=\dfrac{2(a_{1}^2-a_{0}a_{2})}{a_{0}^2}.
\end{equation} 
Analogously, one can also obtain the exact formula for the distribution of the time spent transitioning from chemical states $4$ to $1$:
\begin{equation}
f^{r}(t)=\dfrac{\alpha_{43}\alpha_{32}\alpha_{21}}{(\Omega_{1}-\Omega_{2})(\Omega_{1}-\Omega_{3})}e^{-\Omega_{1}t}+\dfrac{\alpha_{43}\alpha_{32}\alpha_{21}}{(\Omega_{2}-\Omega_{1})(\Omega_{2}-\Omega_{3})}e^{-\Omega_{2}t}+\dfrac{\alpha_{43}\alpha_{32}\alpha_{21}}{(\Omega_{3}-\Omega_{1})(\Omega_{3}-\Omega_{2})}e^{-\Omega_{3}t}.
\end{equation}
Here, $\Omega_{1}$, $\Omega_{2}$ and $\Omega_{3}$ are the solution of the equation
\begin{equation}
\Omega^3-\Omega^2 b_{2}+ \Omega b_{1} - b_{0} =0,
\end{equation}
where
\begin{align}
b_{0}&=\alpha_{43}\alpha_{32}\alpha_{21}, \\
b_{1}&=\alpha_{34}\alpha_{21}+\alpha_{34}\alpha_{23}+\alpha_{32}\alpha_{21}+\alpha_{43}\alpha_{32}+\alpha_{43}\alpha_{21}+\alpha_{43}\alpha_{23},\text{ and} \\
b_{2}&=\alpha_{43}+\alpha_{34}+\alpha_{32}+\alpha_{21}+\alpha_{23}.
\end{align}
Now, solving for the first moment of this distribution gives:
\begin{equation} \label{eqn:4s-tr}
<t_{r}>=\int_{0}^{\infty}t f^{r}(t)dt=\dfrac{b_{1}}{b_{0}}=\dfrac{1}{\alpha_{43}}\biggl[1+\dfrac{\alpha_{34}}{\alpha_{32}}+\dfrac{\alpha_{34}\alpha_{23}}{\alpha_{32}\alpha_{21}}\biggr]+\dfrac{1}{\alpha_{32}}\biggl[1+\dfrac{\alpha_{23}}{\alpha_{21}}\biggr]+\dfrac{1}{\alpha_{21}}.
\end{equation}
and solving for the second moment gives:
\begin{equation}
<t_{r}^2>=\int_{0}^{\infty}t^2 f^{r}(t)dt=\dfrac{2(b_{1}^2-b_{0}b_{2})}{b_{0}^2}.
\end{equation}
%%%%%%%%%%%%%%%%%%%
Assuming that the experimentally observed, fractional population of chemical state i,  $\chi_i$,  represents the equilibrium-state solutions, $P_{\mu}^{eq}$, ratios of $\chi$ can be used to write relationships between several $\alpha$,
\begin{align}
\alpha_{21} = \frac{\chi_1}{\chi_2}\alpha_{12} \hspace{20pt}
\alpha_{32} = \frac{\chi_2}{\chi_3}\alpha_{23} \hspace{20pt}
\alpha_{34} = \frac{\chi_4}{\chi_3}\alpha_{43}. \label{eqn:4s-sub}
\end{align}
The expectation values for the time spent transitioning from chemical states 1 to 4 ($\langle t_p \rangle$), and transitioning from chemical states 4 to 1 ($\langle t_r\rangle$) are assumed to be equivalent to the inverses of the experimentally observed transition rates between the two states of a two-state model ($k_{12}$, and $k_{21}$). Using these expressions and the experimentally observed two-state rates, making substitutions with equations (\ref{eqn:4s-sub}) and rearranging yields the following system of equations,
\begin{align}
\frac{1}{k_{12}} &= 1 \cdot \frac{1}{\alpha_{12}} + C_1 \cdot \frac{1}{\alpha_{23}} + C_2 \cdot \frac{1}{\alpha_{43}}; \hspace{20pt} 
C_1 = \left(\frac{\chi_1}{\chi_2}+1\right) \hspace{20pt} 
C_2 = \left(\frac{\chi_1+\chi_2+\chi_3}{\chi_4}\right) \label{eqn:4s-sys1}\\
\frac{1}{k_{21}} &= C_3 \cdot \frac{1}{\alpha_{12}} + C_4 \cdot \frac{1}{\alpha_{23}} + 1 \cdot \frac{1}{\alpha_{43}}; \hspace{20pt} 
C_3 = \left(\frac{\chi_4+\chi_3+\chi_2}{\chi_1}\right) \hspace{20pt} 
C_4 = \left(\frac{\chi_4+\chi_3}{\chi_2}\right), \label{eqn:4s-sys2}
\end{align}
which has fewer constraints than degrees of freedom. To proceed, we solve the system of equations keeping one degree of freedom independent, moving that term to the left-hand side of the equation, and treating it as part of the constraints.

\subsection*{Independent $\alpha_{12}$}
We solve the system of equations for an independent $\alpha_{12}$ by matrix inversion:

\begin{align}
B=AX \rightarrow
\begin{bmatrix}
\left(\frac{1}{k_{12}} - 1\cdot\frac{1}{\alpha_{12}}\right)\\
\left(\frac{1}{k_{21}} - C_3\cdot\frac{1}{\alpha_{12}}\right)
\end{bmatrix} =
\begin{bmatrix}
C_1 & C_2 \\
C_4 & 1
\end{bmatrix}
\begin{bmatrix}
\frac{1}{\strut\alpha_{23}}\\
\frac{1}{\strut\alpha_{43}}
\end{bmatrix} \Rightarrow
X=A^{-1}B, \label{eqn:4s-matrix} \\
\text{ where } A^{-1} = 
\frac{1}{(C_1\cdot1-C_2C_4)}\cdot
\begin{bmatrix}
1 & -C_2 \\
-C_4 & C_1
\end{bmatrix}, \notag
\end{align}
which yields the following rate constants, 
\begin{align}
\alpha_{12} &= Independent,
\notag\\
\alpha_{21} &= \left(\frac{\chi_1}{\chi_2}\right)\alpha_{12} ,
\notag\\
\alpha_{23} &= \frac{(C_1-C_2C_4)(k_{12}k_{21}\alpha_{12})}{(1\cdot(k_{21}\alpha_{12} - k_{12}k_{21}) - C_2\cdot(k_{12}\alpha_{12} - C_3k_{12}k_{21}))},
\notag\\
\alpha_{32} &= \left(\frac{\chi_2}{\chi_3}\right) \alpha_{23},
\notag\\
\alpha_{34} &= \left(\frac{\chi_4}{\chi_3}\right) \alpha_{43},\text{ and}
\notag\\
\alpha_{43} &= \frac{(C_1-C_2C_4)(k_{12}k_{21}\alpha_{12})}{(-C_4\cdot(k_{21}\alpha_{12} - k_{12}k_{21}) + C_1\cdot(k_{12}\alpha_{12} - C_3k_{12}k_{21}))}. \notag
\end{align}
\subsection*{Independent $\alpha_{23}$}
Similarly, for an independent $\alpha_{23}$,
\begin{align}
\alpha_{12} &= \frac{(1-C_2C_3)(k_{12}k_{21}\alpha_{23})}{(1\cdot (k_{21}\alpha_{23} - C_1k_{12}k_{21})- C_2 \cdot (k_{12}\alpha_{23} - C_4 k_{12} k_{21}))},
\notag\\
\alpha_{21} &= \left(\frac{\chi_1}{\chi_2}\right)\alpha_{12} ,
\notag\\
\alpha_{23} &= Independent,
\notag\\
\alpha_{32} &= \left(\frac{\chi_2}{\chi_3}\right) \alpha_{23},
\notag\\
\alpha_{34} &= \left(\frac{\chi_4}{\chi_3}\right) \alpha_{43},\text{ and}
\notag\\
\alpha_{43} &= \frac{(1-C_2C_3)(k_{12}k_{21}\alpha_{23})}{(-C_3\cdot (k_{21}\alpha_{23} - C_1k_{12}k_{21}) + 1 \cdot (k_{12}\alpha_{23} - C_4 k_{12} k_{21}))}.
\notag
\end{align}
\subsection*{Independent $\alpha_{43}$}
Finally, for an independent $\alpha_{43}$,
\begin{align}
\alpha_{12} &= \frac{(C_4-C_1C_3)(k_{12}k_{21}\alpha_{43})}{(C_4\cdot (k_{21}\alpha_{43} - C_2k_{12}k_{21})-C_1\cdot (k_{12}\alpha_{43} - k_{12}k_{21}))},
\notag\\
\alpha_{21} &= \left(\frac{\chi_1}{\chi_2}\right)\alpha_{12}, 
\notag\\
\alpha_{23} &= \frac{(C_4-C_1C_3)(k_{12}k_{21}\alpha_{43})}{(-C_3\cdot (k_{21}\alpha_{43} - C_2k_{12}k_{21})+1\cdot (k_{12}\alpha_{43} - k_{12}k_{21}))},
\notag\\
\alpha_{32} &= \left(\frac{\chi_2}{\chi_3}\right) \alpha_{23},
\notag\\
\alpha_{34} &= \left(\frac{\chi_4}{\chi_3}\right) \alpha_{43},\text{ and}
\notag\\
\alpha_{43} &= Independent.
\notag
\end{align}

%%%%%%%%%%%%%%%%%%%%%%%%
\section{Three-State Model}
%%%%%%%%%%%%%%%%%%%%%%%
Given the following linear, three-state model,
\begin{equation*}
1 \mathop{\rightleftharpoons}^{\alpha_{12}}_{\alpha_{21}} 2 \mathop{\rightleftharpoons}^{\alpha_{23}}_{\alpha_{32}} 3,
\end{equation*}
the time evolution of the probability, $P_{\mu}(t)$, will be governed by 
\begin{align}
\frac{dP_{1}(t)}{dt}&=-\alpha_{12}P_{1}(t)+\alpha_{21}P_{2}(t), \\
\frac{dP_{2}(t)}{dt}&=\alpha_{12}P_{1}(t)-(\alpha_{21}+\alpha_{23})P_{2}(t)+\alpha_{32}P_{3}(t),\text{ and} \\ 
\frac{dP_{3}(t)}{dt}&=\alpha_{23}P_{2}(t)-\alpha_{32}P_{3}(t).
\end{align}
We calculate the equilibrium-state probabilities   
\begin{align}
P_{1}^{eq}&=\frac{\alpha_{21}\alpha_{32}}{\alpha_{21}\alpha_{32}+\alpha_{12}\alpha_{32}+\alpha_{23}\alpha_{12}} \\
P_{2}^{eq}&=\dfrac{\alpha_{12}\alpha_{32}}{\alpha_{21}\alpha_{32}+\alpha_{12}\alpha_{32}+\alpha_{23}\alpha_{12}} \\
P_{3}^{eq}&=\dfrac{\alpha_{23}\alpha_{12}}{\alpha_{21}\alpha_{32}+\alpha_{12}\alpha_{32}+\alpha_{23}\alpha_{12}}
\end{align}
Then, we also calculate the distribution of the time spent by a molecule transitioning from chemical states $1$ to $3$. For this purpose, we modify the scheme into 
\begin{equation*}
1 \mathop{\rightleftharpoons}^{\alpha_{12}}_{\alpha_{21}} 2 \mathop{\rightarrow}^{\alpha_{23}} 3,
\end{equation*}
and write the master equations according to this new scheme
\begin{align}
\frac{dP_{1}(t)}{dt}&=-\alpha_{12}P_{1}(t)+\alpha_{21}P_{2}(t), \\
\frac{dP_{2}(t)}{dt}&=\alpha_{12}P_{1}(t)-(\alpha_{21}+\alpha_{23})P_{2}(t),\text{ and} \\
\frac{dP_{3}(t)}{dt}&=\alpha_{23}P_{2}(t).
\end{align}
As in the four-state result, we can solve these equations with the Laplace transform method to yield, 
\begin{equation}
f^{p}(t)=\dfrac{\alpha_{12}\alpha_{23}}{\omega_{2}-\omega_{1}}e^{-\omega_{1}t}+\dfrac{\alpha_{12}\alpha_{23}}{\omega_{1}-\omega_{2}}e^{-\omega_{2}t},
\end{equation}
where $\omega_{1}$ and $\omega_{2}$ are the solution of equation
\begin{equation}
\omega^2-\omega(\alpha_{12}+\alpha_{21}+\alpha_{23})+\alpha_{12}\alpha_{23}=0.
\end{equation}
Now, solving for the first moment,
\begin{equation}
<t_{p}>=\dfrac{1}{\alpha_{12}}\biggl[1+\dfrac{\alpha_{21}}{\alpha_{23}}\biggr]+\dfrac{1}{\alpha_{23}},
\end{equation}
and the second moment,
\begin{equation}
<t_{p}^2>=\dfrac{2(c_{1}^2-c_{0})}{c_{0}^2},
\end{equation}
where
\begin{align}
c_{0}&=\alpha_{12}\alpha_{23},\text{ and} \\
c_{1}&=\alpha_{12}+\alpha_{21}+\alpha_{23}.
\end{align}
Similarly, we can also calculate the distribution of the time spent transitioning from chemical states 3 to 1,
\begin{equation}
f^{r}(t)=\dfrac{\alpha_{21}\alpha_{32}}{\Omega_{2}-\Omega_{1}}e^{-\Omega_{1}t}+\dfrac{\alpha_{21}\alpha_{32}}{\Omega_{1}-\Omega_{2}}e^{-\Omega_{2}t},
\end{equation}
where $\Omega_{1}$ and $\Omega_{2}$ are the solution of the equation
\begin{equation}
\Omega^2-\Omega (\alpha_{32}+\alpha_{23}+\alpha_{21})+\alpha_{32}\alpha_{21}=0.
\end{equation}
In this case, we calculate the first moment,
\begin{equation}
<t_{r}>=\dfrac{1}{\alpha_{32}}\biggl[1+\dfrac{\alpha_{23}}{\alpha_{21}}\biggr]+\dfrac{1}{\alpha_{21}},
\end{equation}
and the second moment,
\begin{equation}
<t_{r}^2>=\dfrac{2(d_{1}^2-d_{0})}{d_{0}^2},
\end{equation}
where
\begin{align}
d_{0}&=\alpha_{32}\alpha_{21},\text{ and} \\
d_{1}&=\alpha_{32}+\alpha_{23}+\alpha_{21}.
\end{align}
Assuming that the experimentally observed, fractional population of state i,  $\chi_i$,  represents the equilibrium-state solutions, $P_{\mu}^{eq}$, ratios of $\chi$ can be used to write relationships between several $\alpha$,
\begin{align}
\alpha_{21} = \frac{\chi_1}{\chi_2}\alpha_{12},\text{ and } \hspace{20pt}
\alpha_{23} = \frac{\chi_3}{\chi_2}\alpha_{32}. \label{eqn:3s-sub}
\end{align}
The expectation values for the time spent transitioning from chemical states 1 to 3 ($\langle t_p \rangle$), and transitioning from chemical states 3 to 1 ($\langle t_r\rangle$) are assumed to be equivalent to the inverses of the experimentally observed transition rates between the two final states of a two-state model ($k_{12}$, and $k_{21}$). Using these expressions and experimentally observed rates, substituting Eqn. (\ref{eqn:3s-sub}), and then rearranging yields the following system of equations,
\begin{align}
\langle t_p \rangle = \frac{1}{k_{12}} &= 1\cdot\frac{1}{\alpha_{12}} + C_1\cdot\frac{1}{\alpha_{32}}; \hspace{20pt} C_1 \equiv \left(\frac{\chi_1}{\chi_3}+1\right) \\
\langle t_r \rangle = \frac{1}{k_{21}} &= C_2\cdot\frac{1}{\alpha_{12}} + 1\cdot\frac{1}{\alpha_{32}}; \hspace{20pt} C_2 \equiv \left(\frac{\chi_3 + \chi_2}{\chi_1}\right),
\end{align}
which can be solved as for the four-state model,
\begin{align}
B=AX \rightarrow
\begin{bmatrix}
\frac{1}{\strut k_{12}}\\
\frac{1}{\strut k_{21}}
\end{bmatrix} =
\begin{bmatrix}
1 & C_1 \\
C_2 & 1
\end{bmatrix}
\begin{bmatrix}
\frac{1}{\strut \alpha_{12}}\\
\frac{1}{\strut \alpha_{32}}
\end{bmatrix} \Rightarrow
X = 
\begin{bmatrix}
\frac{1}{\strut k_{12}}\\
\frac{1}{\strut k_{21}}
\end{bmatrix} =
\frac{1}{(1-C_1C_2)}\cdot
\begin{bmatrix}
1 \cdot \frac{1}{\strut k_{12}} - C_2 \cdot \frac{1}{\strut k_{21}}\\
-C_2 \cdot \frac{1}{\strut k_{12}} + 1 \cdot \frac{1}{\strut k_{21}}
\end{bmatrix}, \label{eqn:3s-matrix}
\end{align}
to yield the three-state model rate constants,
\begin{align}
\alpha_{12} &= \frac{(1-C_1C_2)(k_{12}k_{21})}{(1\cdot k_{21} - C_1 \cdot k_{12})}, \notag\\
\alpha_{21} &= \left(\frac{\chi_1}{\chi_2}\right) \alpha_{12},
\notag\\
\alpha_{23} &= \left(\frac{\chi_3}{\chi_2}\right) \alpha_{32},\text{ and}
\notag\\
\alpha_{32} &= \frac{(1-C_1C_2)(k_{12}k_{21})}{(-C_2\cdot k_{21}- 1\cdot k_{12})}.\notag
\end{align}

%%%%%%%%%%%%%%%%%%
\section{smFRET Simulations}
%%%%%%%%%%%%%%%%%%%%

In order to simulate PRE complex dynamics with transient intermediates, we estimated the values of  E$_{\text{FRET}}$ for each PRE complex conformational state. We estimated E$_{\text{FRET}}$ for the L1-tRNA labeling scheme by measuring the distances between the $\beta$-carbon of the threonine at position 202 of the L1 protein of the 50S ribosomal subunit and the sulfur of the thiouridine at position 8 of tRNA$^{fMet}$ from the atomic-resolution, molecular dynamics flexible fitting of the classes of PRE complexes deposited in the Protein Data Bank by Agirrezabala and coworkers (Table \ref{table:efret}).\citep{Agirrezabala2012} While the absolute accuracy of these estimates is likely imprecise, it is reasonable to interpret the relative distances as an informative measure of the relative E$_{\text{FRET}}$ values. As the distances measured ignore any foreshortening due to the space occupied by the fluorophore and its hydrocarbon linker, subsequent analysis compensated by overestimating R$_0$ = 60 \AA. Notably, all the classes measured yielded distinct values of E$_{\text{FRET}}$ except for classes 2 and 4A (MS I and IS1, respectively). Since the distances between the labeling sites on the L1 protein and the P-site tRNA are 78 \AA~(E$_{\text{FRET}} \approx$~0.17) and 81 \AA~(E$_{\text{FRET}} \approx$~0.14) for classes 2 and 4A, respectively, MS I and IS1 are most likely indistinguishable given the SBR of the TIRF-based smFRET measurements used by Fei and coworkers.\citep{Fei2008,Fei2009} As such, we chose to group MS I and IS1 together into state 1, while IS2 corresponded to state 2, and MS II corresponded to state 3. 

\begin{table}%[!HTB] \label{table:efret}
\begin{center}
\begin{tabular}{c | c c}
Class & r(L1 $\rightarrow$ tRNA)$\left(\text{\AA}\right)$ & $E_{\text{FRET}}$, R$_0$ = 60 \AA \\
\hline 
2 & 78 & 0.17 \\
4A & 81 & 0.14 \\
4B & 64 & 0.41 \\
5 & 43 & 0.89 \\
6 & 55 & 0.62 \\
\end{tabular}
\end{center}
\caption{Distances and approximate FRET efficiencies of PRE$_{\text{fM/F}}$ ribosomes from cryo-EM structures.}  \label{table:efret}
\end{table}

From these E$_{\text{FRET}}$ estimates, Markovian transitions along a linear, three-state kinetic scheme were then simulated for 100 state versus time trajectories. Each state versus time trajectory was 50 sec in length, and they were eventually transformed into discrete, E$_{\text{FRET}}$ versus time trajectories, where each data point is the mean E$_{\text{FRET}}$ value during a 50 msec time period. For each state versus time trajectory, the distances between the donor- and acceptor fluorophores in the i$^{th}$ state, $r_i$, were randomized for each time series with a normal distribution, $\mathcal{N}\left( \mu = r_i, \sigma = 2\text{\AA}\right)$. The E$_{\text{FRET}}$ of each state was then calculated as $E_{\text{FRET}} = (1+\left(r/R_0\right)^6)^{-1}$, where $R_0$ is the F\"orster radius (a parameter dependent upon the identity of donor- and acceptor fluorophores, as well their local environments). $R_0$ was randomized for each time series within a reasonable range for the Cy3-Cy5 FRET donor-acceptor pair with a normal distribution, $\mathcal{N}\left( \mu = 60 \text{\AA}, \sigma = 2 \text{\AA} \right)$. The E$_{\text{FRET}}$ versus time trajectory was then discretized by calculating the average value of E$_{\text{FRET}}$ during sequential 50 msec long time periods. Noise was added to the E$_{\text{FRET}}$ versus time trajectories that was normally distributed at each data point with a standard deviation of 0.05 -- a reasonable SBR for data collected on the TIRF microscope used in the smFRET experiments.

In order to model the histogram of the simulated E$_{\text{FRET}}$ versus time trajectories, we used a Gaussian mixture model where each state is modeled to contribute as a normal distribution centered at the respective mean E$_{\text{FRET}}$ value for that state, and is weighted by the equilibrium-state probability for that state (see Appendix B). To account for the simulated heterogeneity in the ensemble of synthetic E$_{\text{FRET}}$ versus time trajectories, the mean E$_{\text{FRET}}$ value of each state was marginalized out by integrating over the joint-probability distribution of the normal distribution of  E$_{\text{FRET}}$ and a beta distribution of the mean E$_{\text{FRET}}$ observed in that state (Fig. 4C) with parameters determined by a maximum likelihood estimate from the exact simulated E$_{\text{FRET}}$ means. For the ``Avg. State'' distribution (\textit{c.f.,} Fig. 4B), the distribution of E$_{\text{FRET}}$ means that was employed was beta distributed with a linear combination of the parameters of the mean E$_{\text{FRET}}$ distributions from states 1 and 2 (Fig. 4C). The standard deviation used for the normal distribution of E$_{\text{FRET}}$ values for each state was taken exactly as the standard deviation used to add noise to the synthetic E$_{\text{FRET}}$ versus time trajectories.

As described in Section \ref{sec:simulation}, this model of the observed E$_{\text{FRET}}$ value histograms is for a temporally-resolved histogram without any blurring present. Performing the same simulation described above, but with a 0.5 msec acquisition time period yields an accurately modeled set of E$_{\text{FRET}}$ versus time trajectories (Fig. 5). Furthermore, analyzing this data with ebFRET yields an accurately estimated number of states, rate constants, distribution of E$_{\text{FRET}}$ means, and noise parameter.

Analysis of the 50 msec time resolution data with ebFRET found the most evidence for a five-state kinetic model--all of which were significantly populated. Since the synthetic data was simulated with a three-state kinetic model, the ebFRET analysis is not consistent with the original simulation. Additionally, the rate constants inferred by ebFRET for a three-state model from the simulated data do not match the original simulation parameters. The rate constants inferred by ebFRET for the two-state model were $k_{GS1\rightarrow GS2} = 1.41 \pm 0.05 s^{-1}$, and $k_{GS2\rightarrow GS1} = 1.41 \pm 0.05 s^{-1}$, but these differ from those learned from the data of Fei and coworkers (\textit{c.f.,} Section \ref{sec:pretranslocation}). This suggests that the original simulation parameters are not consistent with the experimental data from the smFRET study of Fei and coworkers.\citep{Fei2008,Fei2009} Most likely, the discrepancy is due to experimental differences between the smFRET and cryo-EM studies that were compared using the general framework presented here.

%%%%%%%%%%%%%%%%%%%%%%
\begin{acknowledgement} 
%%%%%%%%%%%%%%%%%%%%%%

The authors thank Dr. Jingyi Fei for comments on the manuscript, and Ms. Melissa Thomas and Dr. Wen Li for assistance in the preparation of the figures. This work is supported in part by: the Department of Energy Office of Science Graduate Fellowship Program (DE-AC05-06OR23100) (C.D.K.); the Council of Scientific and Industrial Research (A.K.S.); the Howard Hughes Medical Institute and NIH-NIGMS grants (R01 GM29169 and R01 GM55440) (J.F.); a Burroughs Wellcome Fund CABS Award (CABS 1004856), an NSF CAREER Award (MCB 0644262), an NIH-NIGMS grant (R01 GM084288), an American Cancer Society Research Scholar Grant (RSG GMC-117152), and a Camille Dreyfus Teacher-Scholar Award (R.L.G.); and the Prof. S. Sampath Chair Professorship, and a J.C. Bose National Fellowship (D.C).
\end{acknowledgement} 

%\setlength{\fboxrule}{0 pt}
%\begin{tocentry}
%	\includegraphics{Fig_TOC.pdf}
%\end{tocentry}

%\bibliography{/Users/colin/Documents/Papers/BibTex/dtd_ms}

%%%%%%%%%%%%%%%%%%%%%%%%%%%%%%%
\bibliography{dtd_ms}
%%%%%%%%%%%%%%%%%%%%%%%%%%%%%%%

\end{document}